\documentclass[aps,prl,twocolumn,groupedaddress]{revtex4}
\usepackage[pdftex]{graphicx}
\usepackage{amsmath,amssymb,bm}
\usepackage[mathscr]{eucal}
\usepackage{epsfig}

\newif\ifarxiv
\arxivfalse

\begin		{document}
\def\Nfour	{\mathcal N\,{=}\,4}

\def\Nc		{N_{\rm c}}

\def\start	{{\rm 0}}
\def\half	{{\textstyle \frac 12}}
\def\coeff	#1#2{{\textstyle \frac{#1}{#2}}}

\def\Arxiv      #1 [#2]{\href{http://arxiv.org/abs/#1}{{\tt arXiv:#1 [#2]}}\,}

\title
    {
    Horizon formation and far-from-equilibrium isotropization
    \\in supersymmetric Yang-Mills plasma
    }

\author{Paul~M.~Chesler}
\author{Laurence~G.~Yaffe}

\affiliation
    {Department of Physics, University of Washington, Seattle, WA 98195, USA}

\date{\today}

\begin{abstract}
Using gauge/gravity duality, we study the creation and evolution of anisotropic,
homogeneous strongly coupled $\Nfour$ supersymmetric Yang-Mills plasma.
In the dual gravitational description, this corresponds to horizon formation
in a geometry driven to be anisotropic by a time-dependent change
in boundary conditions.
\end{abstract}


\maketitle

{\it{Introduction}}.---The realization that the quark-gluon plasma (QGP) produced
at RHIC is strongly coupled \cite{Shuryak:2004cy}
has prompted much interest in the study
of strongly coupled non-Abelian plasmas.
Hydrodynamic simulations of heavy ion collisions 
have demonstrated that the QGP produced
at RHIC is well modeled by near-ideal hydrodynamics \cite{Luzum:2008cw},
which is a signature of a strongly coupled system.
The success of hydrodynamic modeling of RHIC collisions suggests that the
produced plasma locally isotropizes over a time scale 
$\tau_{\rm iso} \lesssim 1$ fm/c \cite{Heinz:2004pj}\,.
Understanding the dynamics responsible for such rapid isotropization in
a far-from-equilibrium non-Abelian plasma is a challenge.

Because of the difficulty in studying real time dynamics in QCD at strong coupling,
it is useful to have a toy model where one can study the dynamics of a far from
equilibrium, strongly coupled non-Abelian plasma in a controlled setting.  
One such toy model is strongly coupled $\Nfour$ supersymmetric Yang-Mills theory
(SYM),
where one can use gauge/gravity duality to study the theory
in the limit of large $\Nc$ and large 't Hooft coupling $\lambda$.
This has motivated much work devoted to studying various non-equilibrium
properties of thermal SYM plasma.

We are interested in exploring the physics of isotropization
in far-from-equilibrium non-Abelian plasmas,
in the simplest setting which allows complete theoretical control.
This leads us to focus on the dynamics of homogeneous, but anisotropic, 
states in strongly coupled, large $\Nc$ SYM.
A conceptually simple way to create non-equilibrium states
is to turn on time-dependent background fields coupled to
operators of interest.
To create states in which the stress tensor is anisotropic,
it is natural to consider the response of the theory to a time-dependent
change in the spatial geometry.
For simplicity, we limit attention to geometries which have spatial homogeneity
({\em i.e.}, translation invariance in all spatial directions),
an $O(2)$ rotation invariance, and
a constant spatial volume element.
The most general metric satisfying these conditions may be written as
\begin{equation}
\label{boundarygeometry}
ds^2 = -dt^2 + e^{B_0(t)} \, d \bm x_{\perp}^2 + e^{-2 B_0(t)} \, dx_{||}^2 \,,
\end{equation}
where $\bm x_\perp \equiv \{ x_1, x_2 \}$.

The function $B_0(t)$ describes a time-dependent shear in the geometry;
neglecting (4-dimensional) gravity, $B_0(t)$ is a function
one is free to choose arbitrarily.
We will choose $B_0(t)$ to be asymptotically constant as $t\to \pm\infty$.
We will also choose the initial state to be the SYM vacuum.
A time-dependent geometry will do work on the quantum system.
Consequently, the state in the distant future
will be a non-vacuum state which
(when the geometry is once again static)
will be indistinguishable from a thermal state.
During the evolution, because the metric (\ref{boundarygeometry}) changes in an
anisotropic fashion, the resulting plasma will also be anisotropic with
different pressures ({\em i.e.}, stress tensor eigenvalues)
in the longitudinal ($x_{||}$) and
transverse ($\bm x_{\perp}$) directions.
Spatial translation invariance implies that no hydrodynamic modes can be 
excited.  Therefore, the non-equilibrium plasma produced by the changing metric
(\ref{boundarygeometry}) provides a nice laboratory to study the relaxation
of non-hydrodynamic degrees of freedom in a far from equilibrium setting.
We choose
\begin{equation}
B_0(t) = \half \, c \left [ 1 - \tanh ({t}/{\tau}) \right ].
\label{B0}
\end{equation}
For $c \ne 0$,
this represents a time-dependent rescaling of lengths in transverse directions
relative to those in the longitudinal direction, over a period of order $\tau$.
The lack of any other scale in conformally invariant SYM implies
that the final state energy density will be $\mathcal O(\tau^{-4})$.
Without loss of generality we measure all quantities in units where
$\tau = 1$.

{\it{Gravitational description.}}---
Gauge/gravity duality \cite{Maldacena:1997re}  provides a
gravitational description of large $\Nc$ SYM in which the $5d$
dual geometry is governed by Einstein's equations 
with a cosmological constant.
Einstein's equations imply that the boundary metric
$g_{\mu\nu}^{\rm B}(x)$,
which characterizes the geometry of the spacetime boundary,
is dynamically unconstrained.  
The specification of the boundary metric serves as a boundary condition 
for the $5d$ Einstein equations.
This reflects the fact that the dual field theory (which resides on the boundary)
does not back-react on the boundary geometry,
whereas the boundary geometry can influence the dynamics of the field theory.

We consider a $5d$ geometry which coincides with AdS$_5$ in the distant past.
This geometry is dual to the vacuum of SYM.
A time dependent boundary metric
$g_{\mu\nu}^{\rm B}(x)$
will create gravitational radiation which propagates from the boundary
into the bulk.
This infalling gravitational radiation will lead to the formation
of a horizon
,
which acts as an absorber of gravitational radiation 
--- any radiation which passes through the horizon cannot
escape back to the boundary.
At late times when the boundary geometry is no longer changing,
the bulk geometry outside the horizon will relax and asymptotically
become static.
This is the gravitational description of thermalization in SYM.

Diffeomorphism and translation invariance 
allows one to chose the metric ansatz
\begin{align}
\label{metric}
ds^2 = &-A \, dv^2 +
\Sigma^2 \big [ e^{B} d \bm x_{\perp}^2 + e^{-2 B} dx_{||}^2 \big ] + 2 dr \,dv\,,
\end{align}
where $A$, $B$, and $\Sigma$ are all functions of the radial coordinate $r$ and time $v$ only.
Infalling radial null geodesics have constant values of $v$ (as well as
$\bm x_\perp$ and $x_{||}$).
Outgoing radial null geodesics satisfy ${d r}/{dv } = \frac{1}{2}A$.
At the boundary, located at $r = \infty$,
the coordinate $v$ coincides with the boundary time $t$.
The geometry in the bulk at $v>0$ corresponds to the causal future of $t>0$
on the boundary.  The form of the metric (\ref{metric}) is invariant 
under the residual diffeomorphism  
$
r \rightarrow r + f(v),
$
where $f(v)$ is an arbitrary function.

With a metric of the form (\ref{metric}),
Einstein's equations may be reduced to the following set of differential equations:
\begin{subequations}
\begin{eqnarray}
\label{Seq}
0 &=& \Sigma \, (\dot \Sigma)' + 2 \Sigma' \, \dot \Sigma - 2 \Sigma^2\,,
\\ \label{Beq}
0 &=& \Sigma \, (\dot B)' + {\textstyle \frac{3}{2}}
    \big ( \Sigma' \dot B + B' \, \dot \Sigma \big )\,,
\\  \label{Aeq}
0 &=& A'' + 3 B' \dot B - 12 \Sigma' \, \dot \Sigma/\Sigma^2 + 4\,,
\\  \label{Cr}
0 &= & \ddot \Sigma
    + {\textstyle \frac{1}{2}} \big( \dot B^2 \, \Sigma - A' \, \dot \Sigma \big)\,,
\\ \label{Cv}
0 &=& \Sigma'' + {\textstyle \frac{1}{2}} B'^2 \, \Sigma\,,
\end{eqnarray}
\label{Eeqns}%
\end{subequations}
where, for any function $h(r,v)$,
\begin{equation}
    h' \equiv \partial_r h, \qquad
    \dot h \equiv \partial_v h + {\textstyle \frac{1}{2}} A \, \partial_r h\,.
\end{equation}
Eqs.~(\ref{Cr}) and (\ref{Cv}) are constraint equations;
the radial derivative of Eq.~(\ref{Cr})
and the time derivative of Eq.~(\ref{Cv})
are implied by Eqs.~(\ref{Seq})--(\ref{Aeq}).

The above set of differential equations must be solved subject to
boundary conditions imposed at $r = \infty$.
The requisite condition is simply that the boundary metric
$g_{\mu\nu}^{\rm B}(x)$
coincide with our choice (\ref{boundarygeometry}) of the $4d$ geometry.
In particular, we must have
\begin{equation}
\label{bc}
\lim_{r \rightarrow \infty} \Sigma(r,v)/r \equiv 1\,, \qquad
\lim_{r \rightarrow \infty} B(r,v) \equiv B_0(v)\,.
\end{equation}
One may fix the residual diffeomorphism invariance 
by demanding that
\begin{equation}
\label{gaugefix}
\lim_{r \rightarrow \infty} \left [ A(r,v) - r^2 \right ]/r = 0\,.
\end{equation}
These boundary conditions,
plus initial data satisfying the constraint (\ref{Cv}) 
on some $v={\rm const.}$ slice,
uniquely specify the subsequent evolution of the geometry.

Given a solution to Einstein's equations, the SYM stress tensor is determined
by the near-boundary behavior of the $5d$ metric \cite{deHaro:2000xn}\,.
If $S_{\rm G}$ denotes the gravitational action, then
the SYM stress tensor is given by
$
    T^{\mu \nu}(x) = ({2}/{\sqrt{-g^{\rm B}(x)}})\>
    {\delta S_{\rm G}}/{\delta g^{\rm B}_{\mu \nu}(x)}\,.
$

Near the boundary one may solve
Einstein's equations with a power series expansion in $r$.  
Specifically, $A$, $B$ and $\Sigma$ have asymptotic expansions of the form
\begin{subequations}
\label{series}
\begin{eqnarray}
A(r,v) &= &  \sum_{n=0} \left [\, a_{n}(v) + \alpha_{n}(v) \log r \right ] r^{2-n}\,,
\\
B(r,v) &= & \sum_{n=0} \left [\, b_{n}(v) + \beta_{n}(v) \log r \right ] r^{-n}\,,
\\
\Sigma(r,v) &= & \sum_{n=0} \left [\, s_{n}(v) + \sigma_{n}(v) \log r \right ] r^{1-n}\,.
\label{eq:Sexp}
\end{eqnarray}
\end{subequations}
The boundary conditions (\ref{bc}) and (\ref{gaugefix}) imply that
$b_{0}(v) \equiv B_0(v)$,
$s_{0}(v) \equiv 1$,
$a_{0}(v) \equiv 1$,
and $a_{1}(v) \equiv 0$.
Substituting the above expansions into Einstein's equations
and solving the resulting equations order by order in $r$,
one finds that there is one undetermined coefficient, $b_4(v)$.
All other coefficients are determined by the boundary geometry,
Einstein's equations, and $b_4(v)$
\footnote
{
The coefficient $a_4$ is determined by a first order ordinary
differential equation, which can be obtained
from the condition that the SYM stress tensor be covariantly conserved.
All other coefficients are determined algebraically from
$b_{0}(v)$, $b_{4}(v)$, $a_4(v)$ and their $v$ derivatives. 
}.

By substituting the above series expansions into the variation
of the on-shell gravitational action, one may compute the expectation
value of the stress tensor in terms of the expansion coefficients.  This 
procedure has been carried out in Ref.~\cite{deHaro:2000xn}, so we simply quote
the results.  In terms of the expansion coefficients,
the SYM stress tensor reads
\begin{equation}
T^{\mu}_{\ \nu} = ({N_c^2}/{2 \pi^2}) \>
{\rm diag} (-\mathcal E,\mathcal P_{\perp},\mathcal P_{\perp},\mathcal P_{||} ) \,,
\end{equation}
where
(with $b_0^{(k)} \equiv \partial_v^k b_0$):
\begin{subequations}
\begin{align}
    -\mathcal E ={}&   {\textstyle \frac{3}{4}} a_4 + 
    {\textstyle \frac{1}{256} }
    \Big [3 (b_0^{(1)} )^4+14 (b_0^{(2)} )^2 -4 b_0^{(1)} b_0^{(3)}  \Big],
\\[4pt] \nonumber
    \mathcal P_{\perp} ={}& -{\textstyle \frac{1}{4} } a_4 + b_4 + 
    {\textstyle \frac{1}{768}}\Big [ 21 (b_0^{(1)} )^4 - 468 (b_0^{(1)} )^2 b_0^{(2)} 
\\ & {}
    + 10 (b_0^{(2)} )^2 + 4 b_0^{(1)} b_0^{(3)} + 64 b_0^{(4)} \Big ],
\\ \nonumber 
    \mathcal P_{||}  ={}&  -{\textstyle \frac{1}{4}} a_4 -2 b_4 +
    {\textstyle \frac{1}{768}} \Big [ 21 (b_0^{(1)} )^4 +936 (b_0^{(1)} )^2 b_0^{(2)} 
\\ & {}
    + 10 (b_0^{(2)} )^2 + 4 b_0^{(1)} b_0^{(3)} -128 b_0^{(4)} \Big ].
\end{align}
\end{subequations}%

{\it{Numerics.}}---%
One may solve the Einstein equations (\ref{Seq})--(\ref{Aeq}) for
the time derivatives $\dot\Sigma$, $\dot B$, and $A''$.
Define
\begin{subequations}
\begin{align}
\label{Theta}
    \Theta(r,v) \equiv
    \int_r^{\infty} &dw \left [\Sigma(w,v)^3 - h_1(w,v) \right ] - H_1(r,v)\,,
\\
    \Phi(r,v) \equiv 
    \int_r^{\infty} &dw \left [2 \Theta(w,v) B'(w,v)\, \Sigma(w,v)^{-3/2}\right.
\nonumber \\ &\quad{}
    - h_2(w,v)  \Bigr ] - H_2(r,v)\,,
\label{Phi}
\end{align}
\label{ThetaPhi}%
\end{subequations}
where $H_n$ is an indefinite (radial) integral of $h_n$,
\begin{equation}
\label{heq}
h_n = H'_n \,.
\end{equation}
Then Eqs.~(\ref{Seq})--(\ref{Aeq}) are solved by
\begin{subequations}
\label{reducedeqns}
\begin{eqnarray}
\label{Sigmadot}
\dot \Sigma &=& -2 \Theta \, \Sigma^{-2}, 
\\ \label{Bdot}
\dot B &=& -\coeff {3}{2} \,\Phi \Sigma^{-3/2}\,,
\\ \label{Aeq2}
A'' &=&
- 4 -24  \Theta \,  \Sigma' \Sigma^{-4} + \coeff 92 \Phi B' \, \Sigma^{-3/2}  \,.
\end{eqnarray}
\end{subequations}
The functions $h_n(r,v)$ are not constrained by Einstein's equations ---
their presence inside the integrands of Eq.~(\ref{ThetaPhi})
are compensated by the subtraction of their integrals $H_n(r,v)$.
However, 
since we have chosen the upper limit of integration in Eq.~(\ref{ThetaPhi})
to be $r = \infty$,
the functions $h_n(r,v)$ must be suitably
chosen so that the integrals (\ref{ThetaPhi}) are convergent.
The simplest choice to accomplish this is to set $h_1(r,v)$
equal to the asymptotic expansion of $\Sigma(r,v)^3$ up to order $1/r^k$,
for some $k > 1$,
and to set $h_2(r,v)$ equal to the asymptotic expansion
of $2 \Theta(r,v)B'(r,v)/\Sigma(r,v)^{3/2}$ up to order $1/r^k$.
In our numerical solutions reported below, we use $k \geq 4$.
This choice makes the large $r$ contribution to the integrals in
Eq.~(\ref{ThetaPhi}) quite small.
As the coefficients of the series expansions (\ref{series}) only depend on
$b_{0}(v)$ and $b_{4}(v)$ and their $v$ derivatives,
this choice determines $h_n(r,v)$ in terms of one
unknown function $b_{4}(v)$. 

With the subtraction functions $h_n$ specified by the aforementioned asymptotic 
expansions, integrating Eq. (\ref{heq}) fixes
the compensating integrals $H_n$ 
up to an integration constant which is an arbitrary function of $v$.
Integrating Eq.~(\ref{Aeq2}) for $A(r,v)$ introduces two further
($v$ dependent) constants of integration.
The most direct route for fixing these constants of integration is to 
match the large $r$ behavior of the expressions (\ref{Sigmadot}) and (\ref{Bdot})
and the integrated version of Eq.~(\ref{Aeq2}) to the
corresponding expressions obtained from the series expansions (\ref{series}).
This fixes all integration constants in terms of $b_0$ and~$b_4$.

Our algorithm for solving the initial
value problem with time dependent boundary conditions is as follows.
Given an initial geometry defined by $B(r,v_{\start})$,
one knows $b_4(v_{\start})$.
Integrating the constraint equation (\ref{Cv}), with the fully determined
asymptotic behavior (\ref{eq:Sexp}),
yields
$\Sigma(r,v_{\start})$.
{}From this information, one can
compute $A(r,v_{\start})$ by integrating Eq.~(\ref{Aeq2}).
With $A(r,v_{\start})$,
$B(r,v_{\start})$ and $\Sigma(r,v_{\start})$ known, one can then compute 
the time derivative $\partial_v B(r,v_{\start})$ from Eq.~(\ref{Bdot})
and step forward in time,
\begin{equation}
    B(r,v_{\start} + \Delta v) \approx B(r,v_{\start}) + \partial_v B(r,v_{\start}) \,
    \Delta v \,.
\end{equation}
Repeating the above process using this updated profile of $B$
determines $\Sigma$ and $A$ at time $v_\start+\Delta v$, from which one computes
$\partial_v B$ for the next time step.
For
an initial geometry corresponding to the SYM vacuum,
plus the choice (\ref{B0}) of boundary data, one has
\begin{equation}
\label{initialgeometry}
B(r,-\infty) = c\,, ~~ \Sigma(r,-\infty) = r\,, ~~ A(r,-\infty) = r^2\,.
\end{equation}

An important practical matter is fixing the computation domain in $r$ ---
how far into the bulk does one want to compute the geometry?
If a horizon forms, then one may excise the geometry inside the horizon
as this region is causally disconnected
from the geometry outside the horizon.
Furthermore, one must excise the geometry to avoid
singularities behind horizons \cite{Anninos:1994dj}\,.  
To perform the excision,
one first identifies the location of an
apparent horizon (an outermost marginally trapped surface)
which, if it exists, must lie inside a true horizon \cite{Wald:1984rg}\,.
We have chosen to make the incision slightly inside the location
of the apparent horizon.
For the metric (\ref{metric}), the location $r_h(v)$ of the apparent horizon
is given by $\dot \Sigma(r_h(v),v) = 0$ or, from Eq.~(\ref{Sigmadot}),
$
\Theta(r_h(v),v) = 0\,.
$

\begin{figure}[ht]
\includegraphics[scale=0.2]{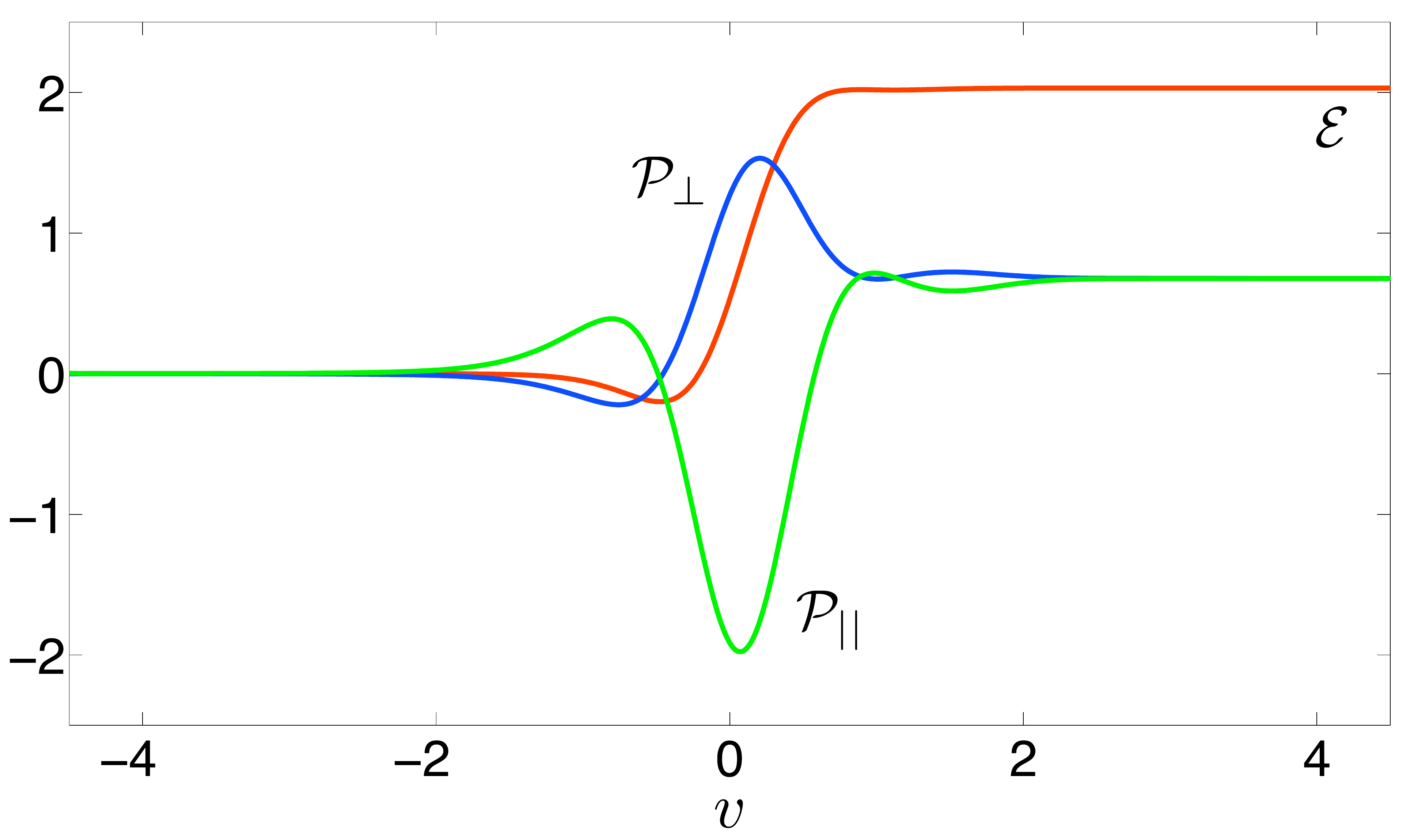}
\caption
    {\label{pressures}
   Energy density, 
   longitudinal
   and transverse pressure,
   all divided by $\Nc^2/2\pi^2$,
   as a function of time for $c = 2$.
   }
\end{figure}

{\it{Results and Discussion.}}---Fig.~\ref{pressures}
shows a plot of the energy density 
and transverse and longitudinal pressures 
produced by the changing boundary geometry (\ref{boundarygeometry}),
with $c = 2$.
These quantities begin at zero in the distant past
when the system is in its vacuum state,
and at late times approach
thermal equilibrium values given by
\begin{equation}
T^{\mu \nu}_{\rm eq} = ({\pi^2 N_c^2 T^4}/{8}) \> {\rm diag} (3,1,1,1),
\end{equation}
where $T$
is the final equilibrium temperature.   
Non-monotonic behavior is seen when the boundary geometry changes most
rapidly around time zero
\footnote
    {
    Nonmonotonicity is unsurprising.
    Late time response for $|c| \ll 1$
    is dominated by
    the lowest quasinormal mode whose frequency is complex,
    $
	\omega_{\rm QNM} = (\pm 9.8 -8.7 i) T
    $
    \cite {Starinets:2002br}.
    }.

\begin{figure}[t]
\includegraphics[scale=0.23]{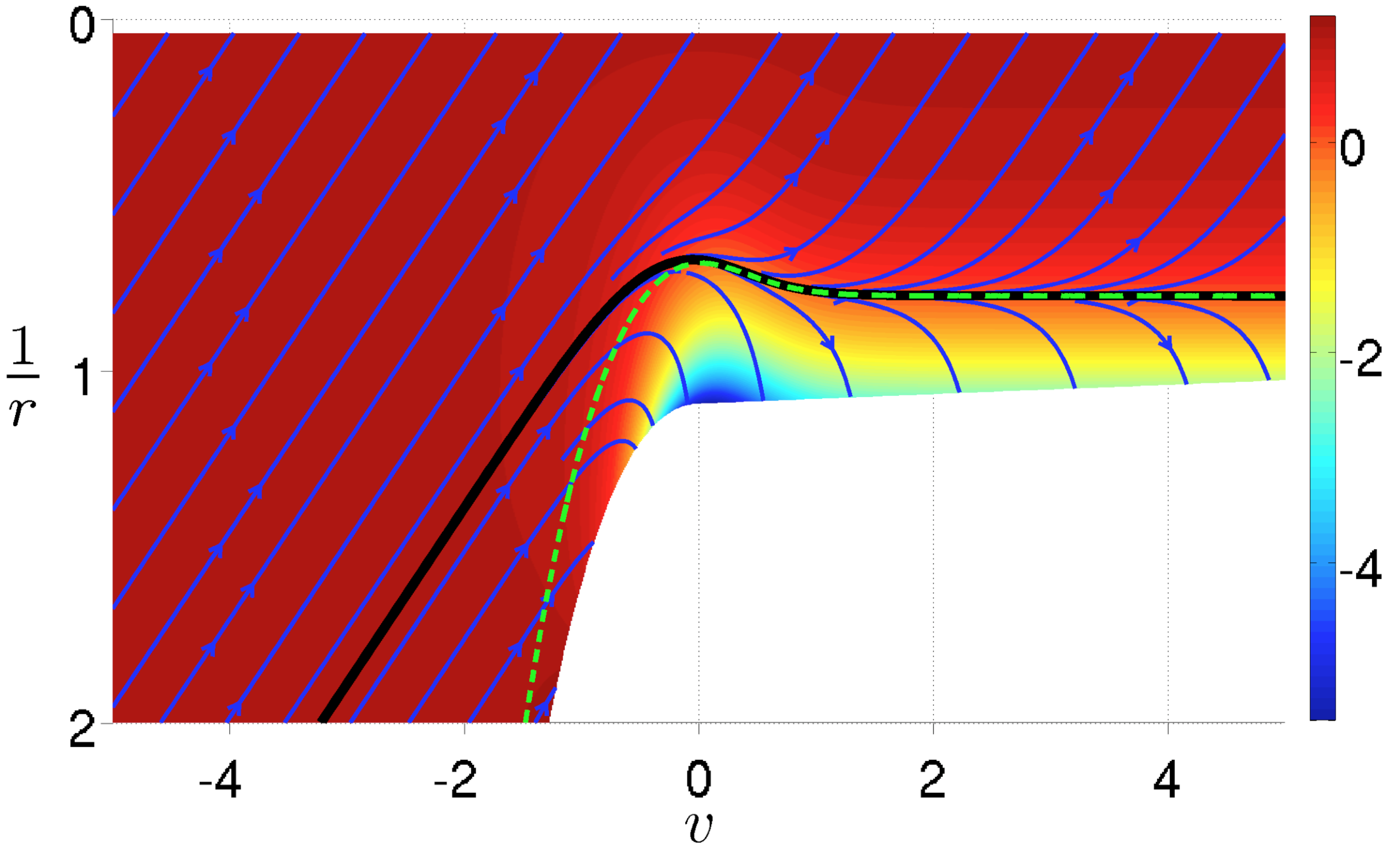}
\caption
  {
  \label{geodesics}
  The congruence of outgoing radial null geodesics.  The surface coloring
  displays $A/r^2$.
  The excised region is 
  beyond the apparent horizon, which is shown by the dashed green line.
  The geodesic shown as a solid black line is the event horizon;
  it separates geodesics which escape to the boundary from those which cannot
  escape.
  }
\end{figure}

Fig.~\ref{geodesics} displays the congruence of outgoing radial null
geodesics, for $c = 2$.  The surface coloring shows $A/r^2$.
In the SYM vacuum ({\em i.e.}, at early times)
this quantity equals $1$,
while at late times $A/r^2 = 1-(r_h/r)^4$.
Excised from the plot is a region of the geometry behind the apparent horizon.
In the SYM vacuum,
outgoing geodesics are given by $1/r + v/2 = {\rm const.}$,
and appear as straight lines in the early part of Fig.~\ref{geodesics}\,.
In the vicinity of $v = 0$, when the boundary geometry is changing rapidly
and producing infalling gravitational radiation,
the geodesic congruence changes dramatically from the zero
temperature form to a finite temperature form.
As is evident from the figure,
at late times some outgoing geodesics do escape to the boundary,
while others fall into the bulk and never escape.
Separating the `escaping' and `plunging' geodesics is one geodesic
which does neither ---
this geodesic, shown as the black line in Fig.~\ref{geodesics},
defines the true event horizon of the geometry.

\begin{figure}[t]
\includegraphics[scale=0.20]{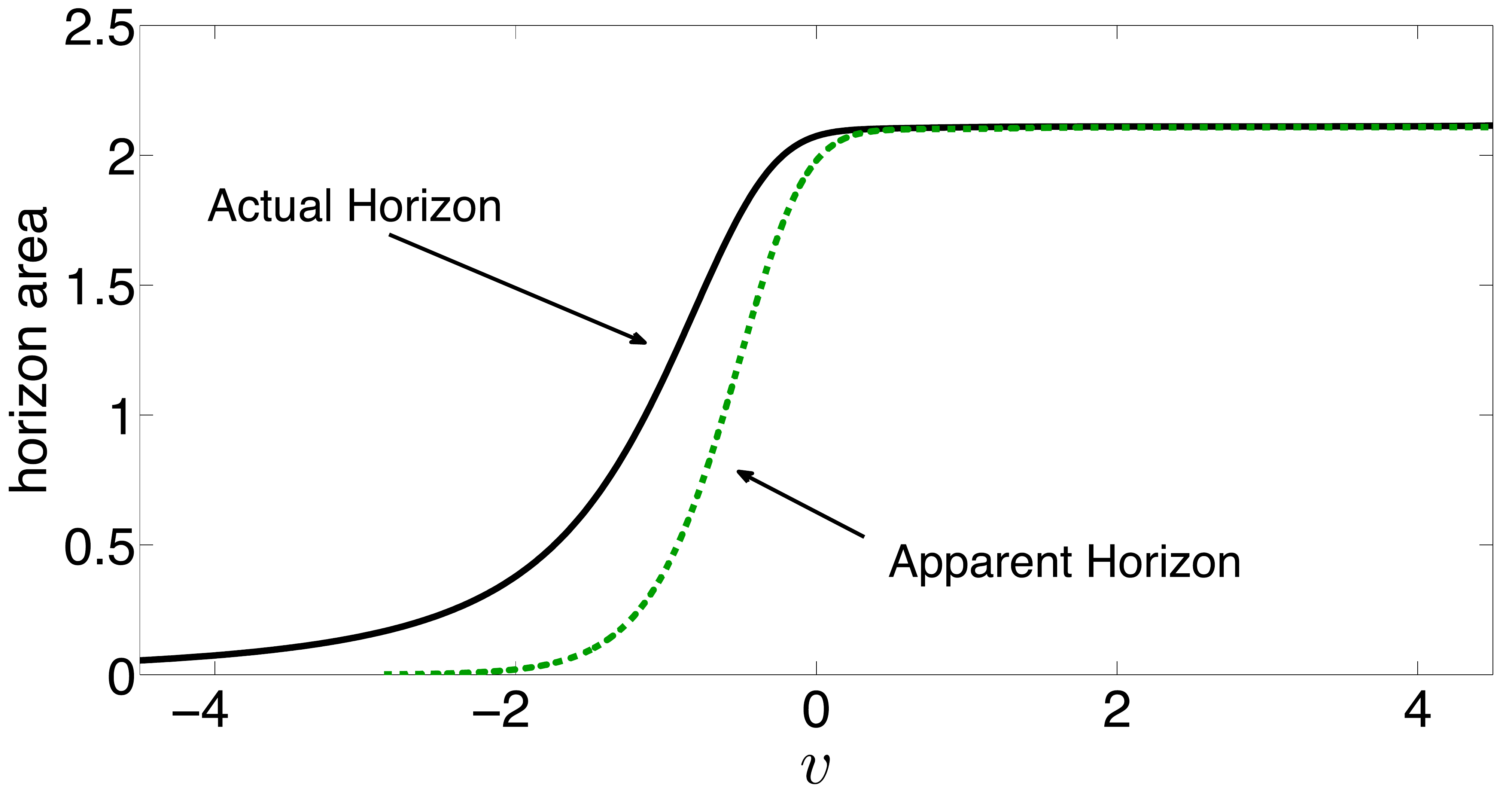}
\caption
    {\label{area}
    Area elements of the true event horizon and the apparent horizon
    as a function of time.
    }
\end{figure}

\begin{table}[t!]
\begin{tabular}{@{\extracolsep{5pt}}c|ccccccc}
$|c|$ & 1 & 1.5 & 2 & 2.5 & 3 & 3.5 & 4
\\
\hline
\hline
$\tau \, T^{\strut}$ & 0.23 & 0.31 & 0.41 & 0.52 & 0.65 & 0.79 & 0.94
\\[2pt]
$\tau_{\rm iso} \, T$ & 0.67 & 0.68 & 0.71 & 0.92 & 1.2 & 1.5 & 1.8
\\[2pt]
$\tau_{\rm iso}/\tau$ & 3.0 & 2.2 & 1.7 & 1.8 & 1.8 & 1.9 & 1.9
\end{tabular}
\caption
    {%
    Final equilibrium temperature $T$
    and isotropization time $\tau_{\rm iso}$ (in units of $T^{-1}$ or $\tau$),
    for various values of $c$.
    The isotropization time
    $\tau_{\rm iso}$ is the time at which the pressures
    deviate from their equilibrium values by less than 10\%.
    }
\label{T1}
\end{table}

Fig.~\ref{area} plots the area of the apparent and 
true event horizons, again for $c = 2$.  
Nearly all growth of the apparent horizon area
occurs in the interval $-2 < v < 0$, during which the boundary geometry is
changing rapidly.
In contrast, the area of the true
horizon grows in the distant past long before the boundary geometry
is significantly perturbed.
This is a reflection of the global nature of event horizons ---
the location of the event horizon depends on the entire history
of the geometry.
It has been argued \cite{Hubeny:2007xt} that it is
the area element of the apparent horizon, pulled back to the boundary along
$v = \rm const.$ infalling null geodesics,
which should be identified with the entropy density (times $4 G_N$)
in the dual field theory.

Table~\ref{T1} shows,
for various values of $c$,
the final equilibrium temperature $T$
and a measure of the isotropization time $\tau_{\rm iso}$.
(These quantities only depend on $|c|$.)
We define $\tau_{\rm iso}$
as the time when the transverse and longitudinal pressures equal
their final values to within $10\%$.
When $|c|\gtrsim 2$, we find that $\tau_{\rm iso} \approx 2 \tau$,
while for $|c| \lesssim 2$, $\tau_{\rm iso} \approx 0.7/T$.
Since
$\tau_{\rm iso}$ is only a few times longer than the time scale $\tau$ over which
the boundary geometry (\ref{boundarygeometry}) is changing,
this measure of isotropization time should best be viewed as an
upper bound on isotropization times associated with plasma dynamics alone.
Nevertheless, it is interesting to note that 
$\tau_{\rm iso} \approx 0.7/T$ corresponds to a time of
$\frac{1}{2}\,$fm/c when $T = 350\,$MeV,
similar to estimates of thermalization times inferred from hydrodynamic modeling
of RHIC collisions \cite{Heinz:2004pj}.

This work was supported in part by the U.S. Department
of Energy under Grant No.~DE-FG02-\-96ER\-40956.
\ifarxiv
L.Y. thanks the Galileo Galilei Institute for Theoretical Physics,
and the Tata Institute for Fundamental Research,
for their hospitality, and the INFN for partial support
while this work was in progress.
We are grateful to
Jim Bardeen, Michal Heller, Rob Myers, Paul Romatschke and Dam Son
for useful discussions.
\bibliographystyle{utphys}
\fi

\bibliography{refs}%
\end{document}